\documentclass[%
 reprint,
showpacs,preprintnumbers,
 amsmath,amssymb,
 aps,
]{revtex4-1}

\usepackage{graphicx}
\usepackage{dcolumn}
\usepackage{bm}
\usepackage{epstopdf}
\usepackage{color}

\begin{document}

\title{Magic and tune-out wavelengths for atomic francium}

\author{U. Dammalapati}
\email{dammalapati.umakanth.a8@tohoku.ac.jp}
\author{K. Harada}
\author{Y. Sakemi}
\affiliation{
 \\Cyclotron and Radioisotope Center (CYRIC), Tohoku University, 6-3 Aramaki-aza Aoba, Aoba-ku, Miyagi 980-8578, Japan
}

\date{\today}

\begin{abstract}
The frequency dependent
polarizabilities of the francium atom are calculated from the available data of energy levels and transition rates. Magic wavelengths for the state insensitive optical dipole trapping are identified from the calculated light shifts of the $7s~^2S_{1/2}$, $7p~^2P_{1/2, 3/2}$ and $8s~^{2}S_{1/2}$ levels of the $7s~^{2}S_{1/2}-7p~^{2}P_{1/2,3/2}$ and $7s~^{2}S_{1/2}-8s~^{2}S_{1/2}$ transitions, respectively.  Wavelengths in the
ultraviolet, visible and near infrared region is identified that
are suitable for cooling and trapping. Magic wavelengths between 600-700~nm and 700-1000~nm region, which are blue and red detuned with the $7s-7p$ and $7s-8s$ transitions are feasible to implement as lasers with sufficient power are available. In addition, we calculated the tune-out wavelengths where the ac polarizability of the ground $7s~^{2}S_{1/2}$ state in francium is zero. These results are beneficial as laser cooled and trapped francium has been in use for
fundamental symmetry investigations like searches for an electron permanent electric dipole moment in an atom and for atomic parity non-conservation.
\end{abstract}

\maketitle

\section{\label{sec:Intr}Introduction}

Radioactive and stable atomic isotopes have been in use as a tool for studies
of nuclear $\beta$-decay and for fundamental symmetry violation
studies such as atomic parity non-conservation (APNC) and searches for a permanent
electric dipole moment (EDM) and other
studies. Traditionally these
experiments were carried out using particles from a long lived parent nuclei and atomic beams. Development of laser cooling and trapping methods produced ultracold atomic samples that resulted in improved manipulation of the atoms and precision measurements~\cite[]{Chu1998,*Phillips1998,*Tannoudji1998}. Specifically, ultracold radioactive samples produced using the above techniques enabled the rare
species to be used more efficiently and minimization of the systematic
effects. This has been demonstrated with $^{82}$Rb~\cite{crane2001},$^{21}$Na~\cite{Scielzo2004},
$^{37m,37}$K~\cite{Gorelov2005}, and $^{225}$Ra~\cite{parker2015}. Recent reviews on the progress and status of these experiments can be found in \cite{Severijns2011,jungmann2013,Behr2014,Roberts2015}.

Additionally, the optical dipole trapping method of ultracold atoms offers many advantages~\cite{Grimm2000}. To mention,  the feasibility of implementation with various configurations, selection of polarization of the laser in addition to the blue and red detuning of frequency with respect to the transition of the atom under investigation. This method of trapping has been found helpful for $\beta$-decay and EDM studies and the advantage of using cold atoms for these studies has been discussed~\cite{Romalis1999,Bijlsma1994,chin2001}. The optical dipole trapping of radioactive $^{82}$Rb~\cite{Feldbaum2007} and $^{225,226}$Ra has been achieved~\cite{parker2012,parker2015}. Further, the `magic
wavelength' concept developed for optical dipole trapping of atoms resulted in unprecedented improvements in ultracold atom
studies~\cite{katori1999,derevianko2011}. Here, our interest is in francium (Fr) atom that is being used for fundamental symmetry violation studies, in particular, search for an electron permanent electron dipole moment~\cite{sakemi2011}. 

Francium, a radioactive heavy alkali metal is one of the widely
used radioactive atoms for fundamental symmetry violation studies. Because of
its high $Z$ and other properties, it is advantageous for studies
of atomic parity violation~\cite{sanguinetti2009,tandecki2014},
searches for an electron permanent electric dipole
moment~\cite{sakemi2011} and nuclear anapole
moment~\cite{gomez2007}. It is also one of the most experimentally
investigated radioactive elements in terms of atomic laser
spectroscopy: measurement of energy
levels~\cite{ekstrom1986,arnold1989,arnold1990,simsarian1996tp,simsarian1999two,grossman2000};
lifetimes~\cite{zhao1997l,simsarian1998,grossman200l,aubin2004,gomez2005}; and
isotope shifts and hyperfine structure~\cite{collister2014}. Laser
cooling and trapping of different Fr isotopes has been
achieved by different groups~\cite{simsarian1996mot,Lu1997,sanguinetti2009,collister2014}. Added to this, there have been many theoretical calculations of energy
levels, dipole matrix elements, lifetimes of states, isotope
shifts and hyperfine structure and the enhancement factors for symmetry violation studies~\cite{dzuba1983en,dzuba1995cal,marinescu1998,biemont1998t,van1998fr,byrnes1999,safronova1999rel,safronova2000high,mukherjee2009,roberts2013a}. This also supports that Fr is a good candidate for the above mentioned
studies. Static and tensor dynamic polarizabilities and magic wavelengths
of naturally abundant alkali elements were
reported~\cite[]{flambaum2008mag,arora2007,safronova2012li,*safronova2013k}. However, there is no information on ac-Stark shifts (light shifts) and magic wavelengths for francium atom. 

In this paper we report calculated ac-Stark shifts of the $7s~^{2}S_{1/2}$ level, $7p~^{2}P_{1/2,3/2}$ levels and the $8s~^{2}S_{1/2}$ level in francium. The light shifts are calculated from the available information of energy levels and transition rates. For the $7s~^{2}S_{1/2}-7p~^{2}P_{1/2,3/2}$, D1 and D2 transitions and the electric dipole forbidden $7s~^{2}S_{1/2}-8s~^{2}S_{1/2}$ transition magic wavelengths are identified for the state insensitive optical dipole trapping, which are useful for different experimental studies. Also, tune-out wavelengths of francium at which the ac polarizability of the ground state becomes zero for the $7s-7p$, $7s-8p$ and $7s-9p$ transitions are presented. 

\section{\label{sec:polarz}Dynamic Polarizabilities}

The formalism of the light shift calculations is similar to that reported in~\cite{dammalapati2012}. In brief, for an atom, the ac-Stark shift or frequency dependent dynamic polarizability of an atomic state
with an angular momentum, $J$ is written as:
\begin{equation}\label{eq1}
    U_i(\omega,q,m_i,r) = -\alpha_i(\omega,q,m_i) I(r)
\end{equation}
where $I(r)$ is the intensity of the laser, $\omega$ is the angular frequency of the radiation, $q$ is the polarization of light and $m_i$ is the magnetic sub-state of the atomic level $i$.
  
The induced dynamic polarizability, $\alpha_i$ can be calculated
from the relation
\begin{equation}\label{eq2}
    \alpha_i(\omega,q,m_i) = \frac{3\pi c^2}{h} \sum_{k,m_k}\frac{A_{ki}
    (2J_k+1)}{\omega^2_{ik}(\omega^2_{ik}-\omega^2)}\left(
  \begin{array}{ c c c }
     J_i&1& J_k  \\
     -m_i&q&m_k \\ \end{array} \right)^2
\end{equation}

where the expression in the large parentheses is the $3J$-symbol, $J_{i}$ and $J_{k}$ are the angular momenta of the initial and final states and $A_{ki}$ is the
transition rate of the transitions with frequencies, $\omega_{ik}$ connected to the state under investigation by electric dipole
selection rules, $c$ is the speed of light and $h$ is the Planck constant. $q=0$ for linear ($\pi$) and $q=\pm 1$ for circular ($\sigma^{\pm}$) polarization of light. For calculations, dipole trap laser power, $p=1~W$ and a beam waist, $w=50~\mu m$ is used. Further, we have also checked our method for rubidium (Rb) atom and the magic wavelengths reported in the literature agree with our values~\cite{arora2007}.

For francium, the $7s~^{2}S_{1/2}-7p~^{2}P_{1/2,3/2}$ transitions are used
for laser cooling and trapping. The
$7s~^{2}S_{1/2}-8s~^{2}S_{1/2}$ transition is of interest
for atomic parity symmetry violation studies.
The spectroscopic data for neutral Fr atom such as energy levels, transitions and transition rates, hyperfine structure constants and isotopes shifts of various Fr isotopes
available from both experimental and theoretical studies has been compiled by
Sansonetti~\cite{sansonetti2007}.
For the calculation of ac-Stark shifts and the corresponding
identification of magic wavelengths and tune-out wavelengths we have taken data of energy levels
and transition rates from Ref.~\cite{sansonetti2007}. Rich
information is available for $^{212}$Fr isotope in terms of energy
levels and transition rates. The calculations carried out in the
present work mainly make use of the $^{212}$Fr isotope data. The
energy levels (cm$^{-1}$) and transition rates ($10^{8}$~s$^{-1}$)
are given in Table.~\ref{wavtranFr}. We have observed two differences in Ref.~\cite{sansonetti2007}. The $8p~^{2}P_{3/2}$ energy level
value given in Table~5 of \cite{sansonetti2007} is higher by 0.004~cm$^{-1}$ from Table~4 (of \cite{sansonetti2007}). In Table~4 of Ref.~\cite{sansonetti2007}, for the
$7p~^{2}P_{3/2}$-$12s~^{2}S_{1/2}$ transition the configuration
term of $12s~^{2}S_{1/2}$ is given as $13s~^{2}S_{1/2}$. Though experimental lifetimes of the $7d~^{2}D_{3/2,5/2}$, $8p~^{2}P_{1/2,3/2}$, $8s~^{2}S_{1/2}$ and $9s~^{2}S_{1/2}$ states are available, they are not used in the present work as there is no information of the branching ratios. For the $7p~^{2}P_{1/2,3/2}-8s~^{2}S_{1/2}$ transitions there is no information of the transition rates, hence not considered.

\begin{table*}[htb]
\begin{center}
\caption{Lower level, upper level, wavenumbers (cm$^{-1}$) and
transition rates (10$^{8}$~s$^{-1}$) in Fr. The number given in square brackets is a power of 10. The wavenumbers and
transition rates are taken from~\cite{sansonetti2007}.}
\begin{tabular}{cccc|cccc}
\hline \hline
  Lower level & Upper level & Wavenumber & Transition rate & Lower level & Upper level & Wavenumber & Transition rate \\
  \hline
 7s~$^{2}S_{1/2}$ &  & 0        &          &   8s~$^{2}S_{1/2}$  &                   & 19739.98   &  \\
 & 7p~$^{2}P_{1/2}$ & 12237.409 & 3.22[-1] &                     & 11p~$^{2}P_{1/2}$ & 30161.07 & 1.29[-3]  \\
 & 7p~$^{2}P_{3/2}$ & 13923.998 & 4.78[-1] &                     & 11p~$^{2}P_{3/2}$ & 30241.60 & 1.30[-3] \\
 & 8p~$^{2}P_{1/2}$ & 23112.96  & 2.64[-2] &                     & 12p~$^{2}P_{1/2}$ & 30840.98 & 7.31[-4] \\
 & 8p~$^{2}P_{3/2}$ & 23658.306 & 2.82[-2] &                     & 12p~$^{2}P_{3/2}$ & 30893.16 & 7.53[-4] \\
 & 9p~$^{2}P_{1/2}$ & 27118.21  & 7.24[-3] &                     & 13p~$^{2}P_{1/2}$ & 31291.72 & 4.66[-4] \\
 & 9p~$^{2}P_{3/2}$ & 27366.20  & 7.52[-3] &                     & 13p~$^{2}P_{3/2}$ & 31327.46 & 4.68[-4]  \\
 & 10p~$^{2}P_{1/2}$ & 29064.18 & 3.02[-3] &                     & 14p~$^{2}P_{1/2}$ & 31605.95 & 3.17[-4] \\
 & 10p~$^{2}P_{3/2}$ & 29198.09 & 3.04[-3] &                     & 14p~$^{2}P_{3/2}$ & 31631.49 & 3.18[-4] \\
 & 11p~$^{2}P_{1/2}$ & 30161.07 & 1.56[-3] &                     & 15p~$^{2}P_{1/2}$ & 31833.76 & 2.28[-4] \\
 & 11p~$^{2}P_{3/2}$ & 30241.60 & 1.56[-3] &                     & 15p~$^{2}P_{3/2}$ & 31852.65 & 2.28[-4] \\
 & 12p~$^{2}P_{1/2}$ & 30840.98 & 9.15[-4] &                     & 16p~$^{2}P_{1/2}$ & 32004.20 & 1.66[-4]  \\
 & 12p~$^{2}P_{3/2}$ & 30893.16 & 9.16[-4] &                     & 16p~$^{2}P_{3/2}$ & 32018.55 & 1.70[-4] \\
 & 13p~$^{2}P_{1/2}$ & 31291.72 & 5.81[-4] &                     & 17p~$^{2}P_{1/2}$ & 32135.03 & 1.29[-4] \\
 & 13p~$^{2}P_{3/2}$ & 31327.46 & 5.81[-4] &                     & 17p~$^{2}P_{3/2}$ & 32146.20 & 1.29[-4] \\
 & 14p~$^{2}P_{1/2}$ & 31605.95 & 3.91[-4] &                     & 18p~$^{2}P_{1/2}$ & 32237.65 & 9.93[-5] \\
 & 14p~$^{2}P_{3/2}$ & 31631.49 & 4.00[-4] &                     & 18p~$^{2}P_{3/2}$ & 32246.51 & 9.92[-5] \\
 & 15p~$^{2}P_{1/2}$ & 31833.76 & 2.81[-4] &                     & 19p~$^{2}P_{1/2}$ & 32319.63 & 7.99[-5] \\
 & 15p~$^{2}P_{3/2}$ & 31852.65 & 2.81[-4] &                     & 19p~$^{2}P_{3/2}$ & 32326.77 & 7.98[-5] \\
 & 16p~$^{2}P_{1/2}$ & 32004.20 & 2.11[-4] &                     & 20p~$^{2}P_{1/2}$ & 32386.15 & 6.56[-5] \\
 & 16p~$^{2}P_{3/2}$ & 32018.55 & 2.10[-4] &                     & 20p~$^{2}P_{3/2}$ & 32391.99 & 6.55[-5]\\
 & 17p~$^{2}P_{1/2}$ & 32135.03 & 1.57[-4] &   7p~$^{2}P_{3/2}$  &                   & 13923.998 & \\
 & 17p~$^{2}P_{3/2}$ & 32146.20 & 1.57[-4] &                     & 7d~$^{2}D_{3/2}$   & 24244.03  & 2.09[-2] \\
 & 18p~$^{2}P_{1/2}$ & 32237.65 & 1.23[-4] &                     & 7d~$^{2}D_{5/2}$   & 24332.93  & 1.29[-1]  \\
 & 18p~$^{2}P_{3/2}$ & 32246.51 & 1.23[-4] &                     & 9s~$^{2}S_{1/2}$   & 25671.00  & 3.66[-2]  \\
 & 19p~$^{2}P_{1/2}$ & 32319.63 & 9.82[-5] &                     & 8d~$^{2}D_{3/2}$   & 27600.657 & 1.13[-2] \\
 & 19p~$^{2}P_{3/2}$ & 32326.77 & 1.00[-4] &                     & 8d~$^{2}D_{5/2}$   & 27645.373 & 6.93[-2] \\
 & 20p~$^{2}P_{1/2}$ & 32386.15 & 8.01[-5] &                     & 10s~$^{2}S_{1/2}$  & 28310.617 & 1.90[-2]\\
 & 20p~$^{2}P_{3/2}$ & 32391.99 & 8.00[-5] &                     & 9d~$^{2}D_{3/2}$   & 29316.497 & 6.41[-3] \\

  7p~$^{2}P_{1/2}$ &   & 12237.409   & &                         & 9d~$^{2}D_{5/2}$   & 29341.817 & 3.91[-2] \\
 & 7d~$^{2}D_{3/2}$  & 24244.03  & 1.66[-1] &                    & 11s~$^{2}S_{1/2}$  & 29718.909 & 1.10[-2] \\
 & 9s~$^{2}S_{1/2}$  & 25671.00  & 2.75[-2] &                    & 10d~$^{2}D_{3/2}$  & 30309.962 & 3.99[-3] \\
 & 8d~$^{2}D_{3/2}$  & 27600.657 & 8.04[-2] &                    & 10d~$^{2}D_{5/2}$  & 30325.605 & 3.56[-2] \\
 & 10s~$^{2}S_{1/2}$ & 28310.617 & 1.33[-2] &                    & 12s~$^{2}S_{1/2}$  & 30559.504 & 6.84[-3] \\
 & 9d~$^{2}D_{3/2}$  & 29316.497 & 4.44[-2] &                    & 11d~$^{2}D_{3/2}$  & 30936.325 & 2.65[-3] \\
 & 11s~$^{2}S_{1/2}$ & 29718.909 & 7.38[-3] &                    & 11d~$^{2}D_{5/2}$  & 30946.643 & 1.58[-2] \\
 & 10d~$^{2}D_{3/2}$ & 30309.962 & 2.67[-2] &                    & 13s~$^{2}S_{1/2}$  & 31101.539 & 4.60[-3] \\
 & 12s~$^{2}S_{1/2}$ & 30559.504 & 4.56[-3] &                    & 12d~$^{2}D_{3/2}$  & 31356.506 & 1.80[-3]\\
 & 11d~$^{2}D_{3/2}$ & 30936.325 & 1.76[-2] &                    & 12d~$^{2}D_{5/2}$  & 31363.655 & 1.09[-2]\\
 & 13s~$^{2}S_{1/2}$ & 31101.539 & 3.05[-3] &                    & 14s~$^{2}S_{1/2}$  & 31471.465 & 3.25[-3] \\
 & 12d~$^{2}D_{3/2}$ & 31356.506 & 1.19[-2] &                    & 13d~$^{2}D_{3/2}$  & 31652.000 & 1.32[-3] \\
 & 14s~$^{2}S_{1/2}$ & 31471.465 & 2.14[-3] &                    & 13d~$^{2}D_{5/2}$  & 31657.155 & 2.35[-2] \\
 & 13d~$^{2}D_{3/2}$ & 31652.000 & 8.48[-3] &                    & 15s~$^{2}S_{1/2}$  & 31735.182 & 2.37[-3] \\
 & 15s~$^{2}S_{1/2}$ & 31735.182 & 1.56[-3] &                    & 14d~$^{2}D_{3/2}$  & 31867.682 & 9.77[-4] \\
 & 14d~$^{2}D_{3/2}$ & 31867.682 & 6.28[-3] &                    & 14d~$^{2}D_{5/2}$  & 31871.514 & 1.74[-2] \\
 & 16s~$^{2}S_{1/2}$ & 31929.789 & 1.15[-3] &                    & 16s~$^{2}S_{1/2}$  & 31929.789 & 1.75[-3] \\
 & 15d~$^{2}D_{3/2}$ & 32029.909 & 4.84[-3] &                    & 15d~$^{2}D_{3/2}$  & 32029.909 & 7.37[-4]  \\
 & 17s~$^{2}S_{1/2}$ & 32077.492 & 8.88[-4] &                    & 15d~$^{2}D_{5/2}$  & 32032.821 & 4.48[-3] \\
 & 16d~$^{2}D_{3/2}$ & 32154.979 & 3.81[-3] &                    & 17s~$^{2}S_{1/2}$  & 32077.492 & 1.35[-3] \\
 & 18s~$^{2}S_{1/2}$ & 32192.251 & 6.97[-4] &                    & 16d~$^{2}D_{3/2}$  & 32154.979 & 5.80[-4] \\
 & 17d~$^{2}D_{3/2}$ & 32253.449 & 2.98[-3] &                    & 16d~$^{2}D_{5/2}$  & 32157.274 & 3.45[-3] \\
 & 19s~$^{2}S_{1/2}$ & 32283.180 & 5.59[-4] &                    & 18s~$^{2}S_{1/2}$  & 32192.251 & 1.06[-3] \\
 & 18d~$^{2}D_{3/2}$ & 32332.354 & 2.39[-3] &                    & 17d~$^{2}D_{3/2}$  & 32253.449 & 4.55[-4] \\
 & 20s~$^{2}S_{1/2}$ & 32356.444 & 4.57[-4] &                    & 17d~$^{2}D_{5/2}$  & 32255.275 & 2.77[-3] \\
 & 19d~$^{2}D_{3/2}$ & 32396.552 & 2.00[-3] &                    & 19s~$^{2}S_{1/2}$  & 32283.180 & 8.73[-4] \\
 & 20d~$^{2}D_{3/2}$ & 32449.483 & 1.67[-3] &                    & 18d~$^{2}D_{3/2}$  & 32332.354 & 3.73[-4] \\
 &                  &          &          &                      & 18d~$^{2}D_{5/2}$  & 32333.827 & 6.66[-3] \\
 &                  &          &          &                      & 20s~$^{2}S_{1/2}$  & 32356.444 & 6.99[-4] \\
 &                  &          &          &                      & 19d~$^{2}D_{3/2}$  & 32396.552 & 3.13[-4] \\
 &                   &          &          &                     & 19d~$^{2}D_{5/2}$  & 32397.761 & 1.86[-3] \\
 &                   &          &          &                     & 20d~$^{2}D_{3/2}$  & 32449.483 & 2.62[-4] \\
 &                   &          &          &                     & 20d~$^{2}D_{5/2}$  & 32450.488 & 1.55[-3] \\

\hline \hline
\end{tabular}
\label{wavtranFr}
\end{center}
\end{table*}

\subsection{$7s~^{2}S_{1/2}-7p~^{2}P_{3/2}$ transition}
\begin{figure}[htb]
\includegraphics[width=82 mm,angle=0]{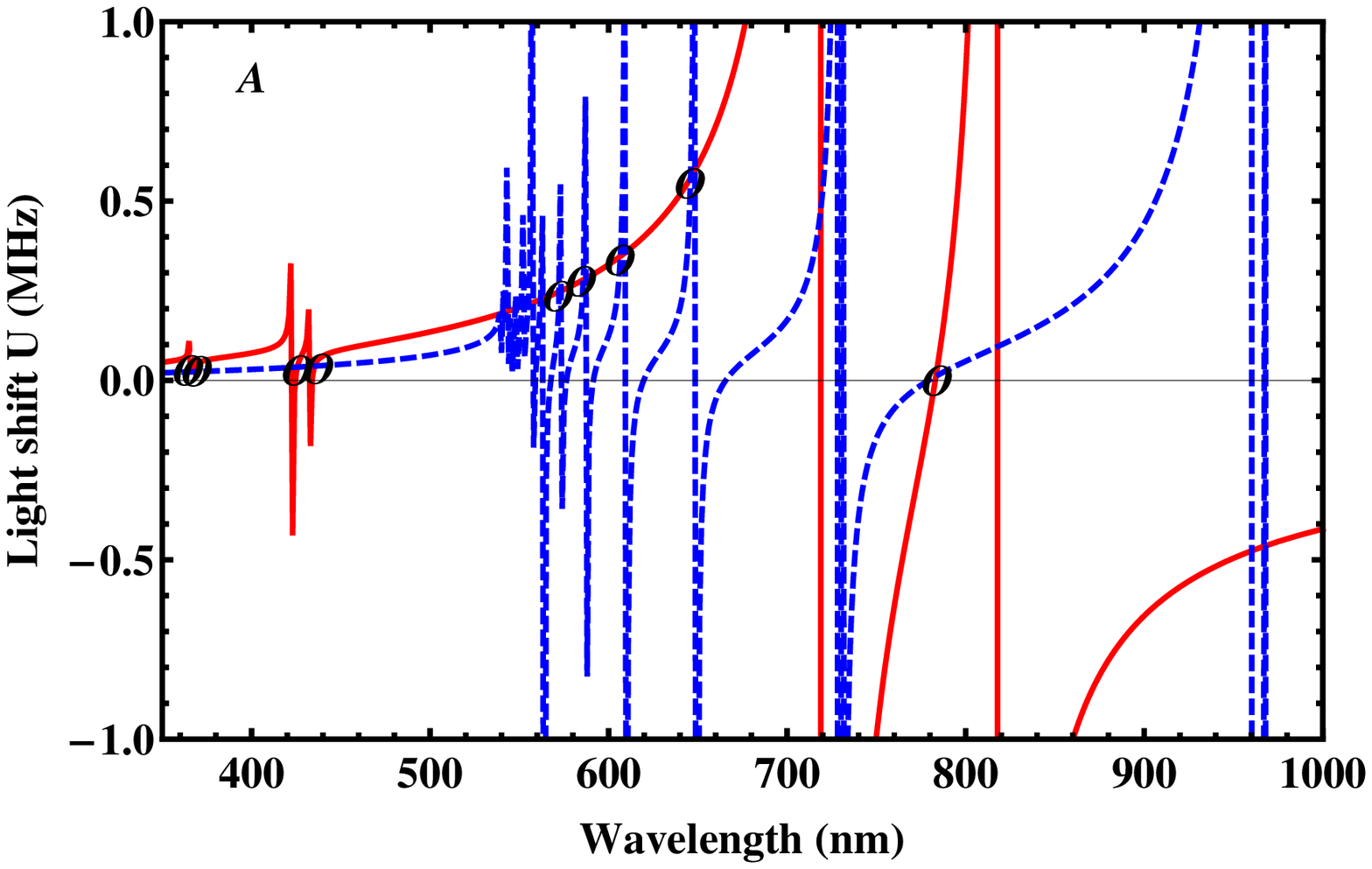}
\includegraphics[width=82 mm,angle=0]{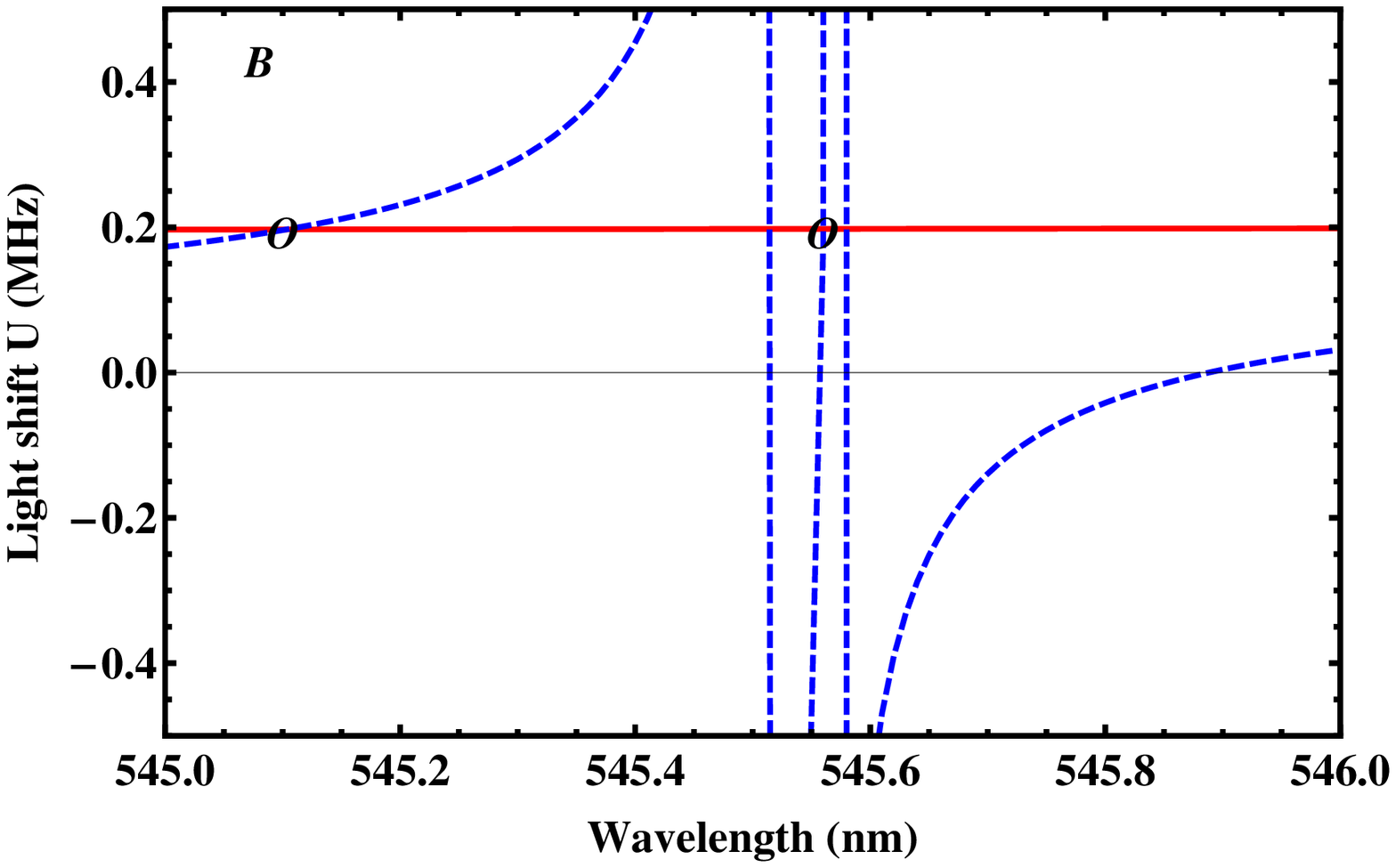}
\caption{\label{7s-7p32m1} (Color online) \textbf{A}) Light shifts from 350~nm to 1000~nm of the ground $7s~^{2}S_{1/2}$ level (red solid line) for $m_J=1/2$ state and the excited $7p~^{2}P_{3/2}$ level (blue dashed line) for $m_J=3/2$ state corresponding to the D2 line in francium are shown. \textbf{B}) Light shifts of the ground $7s~^{2}S_{1/2}$ level ($m_J=1/2$, red solid line) and the excited $7s~^{2}P_{3/2}$ level ($m_J=3/2$, blue dashed line) between 545 and 546~nm range. Similar lines exist in the 540-580~nm range. The positions of the magic wavelengths are denoted with slanted circles (black).}
\end{figure}

The ground state of the neutral Fr atom with the configuration $7s~^{2}S_{1/2}$ and the excited $7p~^{2}P_{3/2}$ level corresponds to the D2 transition at 718.185~nm (vacuum). The dynamic polarizabilities are calculated for the $7s~^{2}S_{1/2}$ level considering the $np_{1/2,3/2}~(n=7-20)$ levels of all electric dipole allowed transitions. In Fig.~\ref{7s-7p32m1} the red solid lines are the ac-Stark shift spectra of the $7s~^{2}S_{1/2}$ ($m_J=1/2$) state. Similarly, the dynamic polarizabilities of the excited $7p~^{2}P_{3/2}$ level are calculated considering the transitions connected from this level to the $ns_{1/2}~(n=9-20)$ and $nd_{3/2,5/2}~(n=7-20)$ states. The light shifts are shown in Fig.~\ref{7s-7p32m1}A as blue dashed lines for the $7p~^{2}P_{3/2}$ ($m_J=3/2$) state as a function of detuning of the perturbing laser for linear polarization of the light. For the $7s~^{2}S_{1/2}$ level, the dominant contribution to the polarizability is from the $7p~^{2}P_{1/2}$, $7p~^{2}P_{3/2}$, $8p~^{2}P_{1/2}$, $8p~^{2}P_{3/2}$ states and to some extent $9p~^{2}P_{1/2}$ and $9p~^{2}P_{3/2}$ levels. Contribution from higher excited level transitions is negligible and are not shown in Fig.~\ref{7s-7p32m1}A. Also, the availability of laser sources around 300~nm with sufficient power to implement a dipole trap is less feasible.  The identified magic wavelengths are denoted as slanted black circles. 

Magic wavelengths in the ultraviolet (UV), visible (VIS) and near infrared (NIR) region are identified. Two magic wavelengths are identified for the $9p~^{2}P_{1/2,3/2}$ states at 365.99~nm and 369.14~nm and two at 425.63~nm and 435.26~nm corresponding to the $8p~^{2}P_{1/2,3/2}$ states, respectively. In the 500-600~nm wavelength region eighteen (18) magic wavelengths are located (all the wavelengths are not marked in Fig.~\ref{7s-7p32m1}A. We have marked only the wavelengths that are well separated for clarity). Of them, two magic wavelengths between 545~nm and 546~nm are shown in Fig.~\ref{7s-7p32m1}B, which are very close. Similarly, four magic wavelengths in the red region, 600-700~nm and five wavelengths in the near infrared region at 729.73~nm, 731.77~nm, 783.35~nm, 967.19~nm and 1111.7~nm, respectively are identified. Lasers with sufficient power to implement an optical dipole trap at magic wavelengths in the 600-1000~nm region are available. Another advantage is that the red wavelengths correspond to the blue detuning with respect to the resonance frequency of the D2 transition of francium. The wavelength at 783.35~nm is close to the tune-out wavelength 782.96~nm (See Table~\ref{tuneoutfr}). However, the photon scattering rate is 2.7~s$^{-1}$ for the considered dipole trap parameters, which will limit the coherence time of the trapped atoms. We have also checked magic wavelength values using the transition rates from the experimental lifetimes of the $7p~^{2}P_{1/2,3/2}$ states. The observed difference is below 0.1~nm. In addition, we have also calculated the light shifts for the circular ($\sigma$) polarization of the laser light and designated the magic wavelengths. All the magic wavelengths obtained for linear and circular polarizations of the dipole trapping laser are given in Table~\ref{magicwavfr1}. The uncertainty of the identified magic wavelengths lies between 0.03~nm and 0.06~nm, which is from the wavelengths of the transitions and in identification of the position.

\begin{figure}[htb]
\includegraphics[width=82 mm,angle=0]{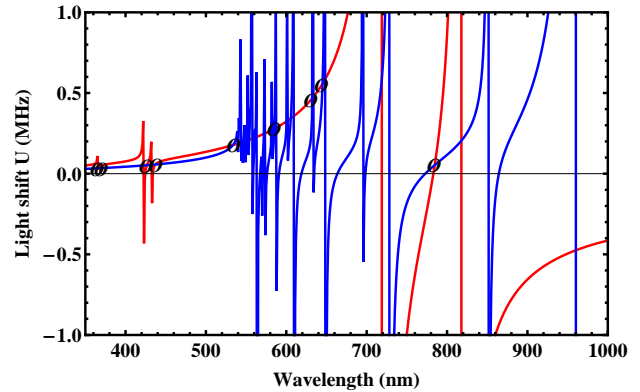}
\caption{\label{7s-7p32m2} (Color online) Light shifts of the $7s~^{2}S_{1/2}$ ($m_J=1/2$) ground state (red solid line) and the $7p~^{2}P_{3/2}$ ($m_J=1/2$) level (blue solid line) are shown for the linear polarization of the dipole trapping laser. Between 440~nm and 510~nm there are two magic wavelengths as the light shift of the p-level crosses at two points. The position of the magic wavelengths is marked with a slanted black circle.}
\end{figure}    

Light shifts are also calculated for the $7s~^{2}S_{1/2}$ ($m_J=1/2$) state and the $7p~^{2}P_{3/2}$ ($m_J=1/2$) state of the $7s~^{2}S_{1/2}-7p~^{2}P_{3/2}$ transition. The obtained light shifts for the linear polarization of the laser light are shown in Fig.~\ref{7s-7p32m2}. The red solid line corresponds to the $7s~^{2}S_{1/2}$ level, the blue dotted line corresponds to the $7p~^{2}P_{3/2}$ level and magic wavelengths are marked with slanted black circles. In total thirty six magic wavelengths are identified between 360~nm and 1200~nm extending from UV region to NIR infrared region: four magic wavelengths in 350-500~nm region corresponding to the $8p$ and $9p$ levels; twenty in the range of 500-600~nm; six in 600-700~nm region; and six between 700-1200~nm. The light shifts and magic wavelengths calculated for circular polarization of the dipole trapping laser are given in Table~\ref{magicwavfr1}.

Another parameter that needs to be considered for optical dipole trapping of atoms is the heating rate. The photon scattering rate from the dipole trap laser light gives rise to heating of the atoms and consequent loss from the trap. We have estimated the photon scattering rates at different magic wavelengths. For example, a dipole trap laser at 645~nm scatter photons at 1.5~s$^{-1}$ and a laser at 1112~nm scatters about 0.15~s$^{-1}$.  

The operation of the dipole trap at magic wavelengths minimizes all the systematic affects due to light shifts. There has been a proposal for measuring an electron EDM using cesium (Cs) atoms in optical lattices and the advantage of using blue tuned dipole trap with linear polarization has been discussed~\cite{chin2001}. In our experiment, we plan to implement an optical dipole trap laser using the linear polarization of light. A detailed evaluation of all the systematics specific to Fr EDM experiment is beyond the scope of this paper.

\subsection{$7s~^{2}S_{1/2}-7p~^{2}P_{1/2}$ transition}
\begin{figure}[htb]
\includegraphics[width=82 mm,angle=0]{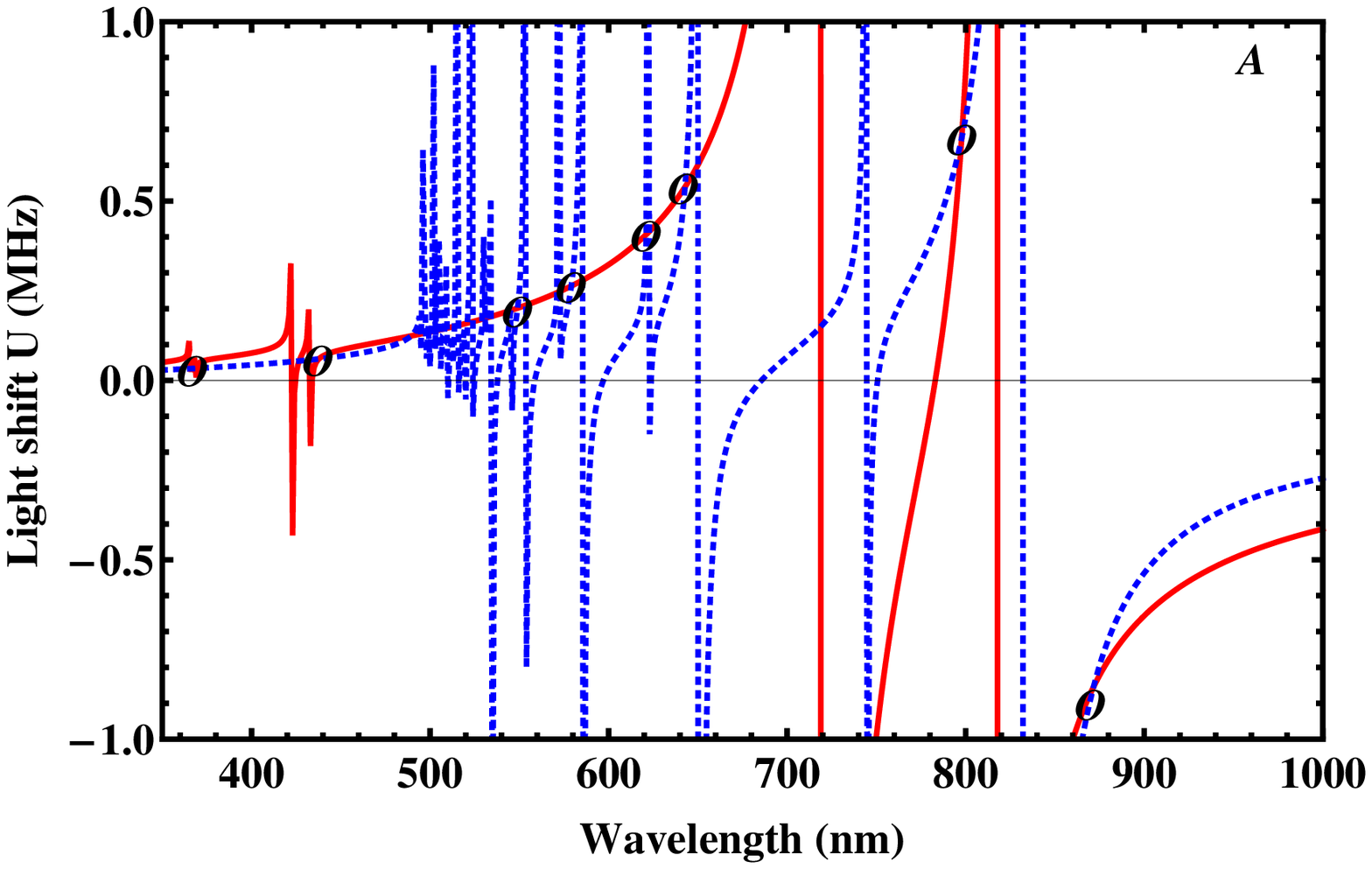}
\includegraphics[width=82 mm,angle=0]{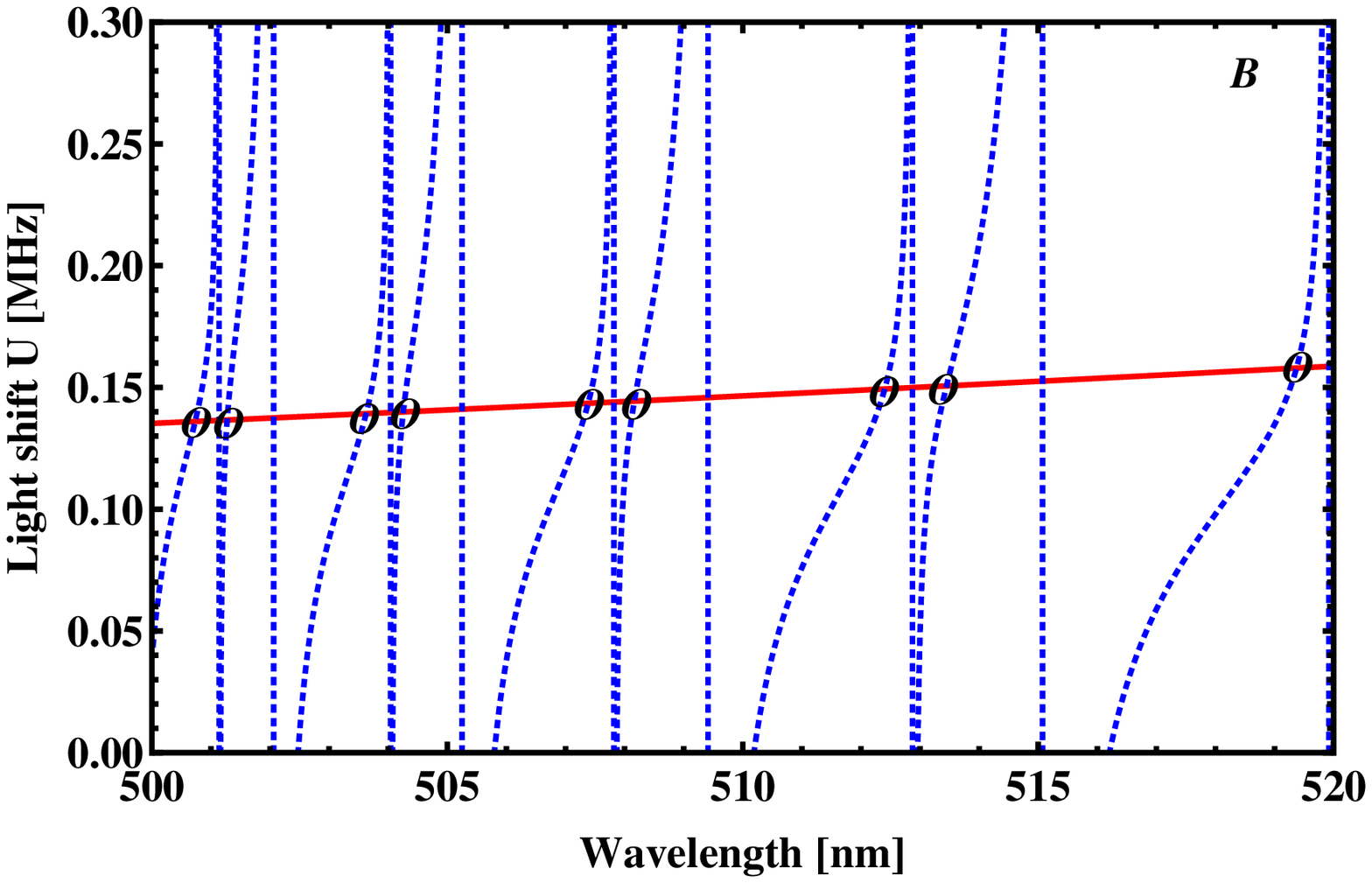}
\caption{\label{7s-7p12} (Color online) \textbf{A}) Light shifts of the ground $7s~^{2}S_{1/2}$ ($m_J=1/2$) state (red solid line) and the excited $7p~^{2}P_{1/2}$ ($m_J=1/2$) state (blue dotted line) corresponding to the D1 transition of the francium atom for linearly polarized light in the wavelength range 300-1000~nm. Of all the magic wavelengths identified, wavelengths in the red region at 621.11~nm, 642.85~nm and near infrared region at 797.75~nm and 871.62~nm are shown. \textbf{B}) The lights shifts of the $7s~^{2}S_{1/2}$ level and the $7p~^{2}P_{1/2}$ level for the 500-520~nm range are shown for the linear polarization of the laser. Positions of the ten magic wavelengths not shown in Fig.~\ref{7s-7p12}A are also indicated with slanted circles (black).}
\end{figure}

The $7s~^{2}S_{1/2}-7p~^{2}P_{1/2}$ transition, also called the D1 line in francium has a vacuum wavelength of 817.166~nm. It has been used as a repumping transition for laser cooling and trapping of francium isotopes. Light shift calculations for this transition are carried out considering both linear and circular polarizations of the optical dipole trapping laser. Shown in Fig.~\ref{7s-7p12}A are the light shifts of the ground $7s~^{2}S_{1/2}$ ($m_J=1/2$) level (red solid line) and the excited $7p~^{2}P_{1/2}$ level for the $m_J=1/2$ state (blue dashed line) from 350~nm to 1000~nm. The polarizabilities  are calculated considering the $nd_{3/2}~(n=7-11)$ and $ns_{1/2}~(n=9-20)$ states. In this transition, we have identified total thirty one (31) magic wavelengths spread from ultraviolet to near infrared region for linear polarization of the laser light. In the 350-500~nm region ten magic wavelengths are identified. Sixteen magic wavelengths in the 500-600~nm region are identified, which are given in Table~\ref{magicwavfr1}. In the red wavelength region, 600-700~nm, two magic wavelengths at 621.1~nm and 642.8~nm are identified and are shown in Fig.~\ref{7s-7p12}A. In the near infrared region, magic wavelengths at 745.40~nm, 797.75~nm and 871.62~nm are identified. In Fig.~\ref{7s-7p12}B, ten magic wavelengths that are very close and not shown in Fig.~\ref{7s-7p12}A between 500~nm and 520~nm are presented. Similar calculations for circular polarization of the laser light are performed. A magic wavelength at 1116.2~nm is identified. The magic wavelengths for the $m_J=-1/2$ state and the $m_J=+1/2$ state are also given in Table~\ref{magicwavfr1}. 

\subsection{$7s~^{2}S_{1/2}~-~8s~^{2}S_{1/2}$ transition}
\begin{figure}[htb]
\includegraphics[width=82 mm,angle=0]{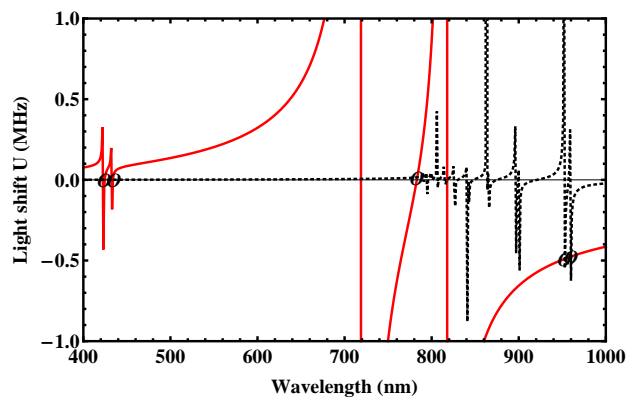}
\caption{\label{7s-8s} (Color online) Light shifts of the $7s~^{2}S_{1/2}$ ground state (red solid line) and the $8s~^{2}S_{1/2}$ level (black dotted line) for 350-1000~nm region are shown for the linear polarization of light. The identified five magic wavelengths are marked with slanted (black) circles.}
\end{figure}

The forbidden $7s~^{2}S_{1/2}~-~8s~^{2}S_{1/2}$ transition in francium is of interest for atomic parity non-conservation studies. Two photon spectroscopy and lifetime of the $8s~^{2}S_{1/2}$ state has been reported~\cite{simsarian1999two,gomez2005}. We have calculated the magic wavelengths for the state insensitive optical dipole trapping of laser cooled francium atoms. To calculate the light shifts, for the $7s~^{2}S_{1/2}$ level, we have considered energy levels and transition rates of the $np~^{2}P_{1/2,3/2}~(n=7-20)$ states and for the $8s~^{2}S_{1/2}$ level $np~^{2}P_{1/2,3/2}~(n=11-20)$ states. The obtained light shifts for the two levels are shown in Fig.~\ref{7s-8s} for the linear polarization of the trapping laser. Three magic wavelengths at 783.46~nm, 953.09~nm and 960.10~nm in the near infrared region of 700-1000~nm are identified. In particular, the magic wavelength at 783.46~nm for the linear polarization of the light is 0.5~nm above the tune-out wavelength (see sec.\ref{sec:tuneout}) of the ground state for the $7s-7p$ transitions. Similarly for circular polarization of the trapping laser, five magic wavelengths for right (and left) circularity of the laser light are identified. The magic wavelengths at 953.09~nm ($\pi$) and 954.01 nm ($\sigma$) could be used as these wavelengths are about 450~nm away from the two photon transition and tuning the laser wavelength by 1~nm is possible. The photon scattering rate is 0.3~s$^{-1}$. One can also check for the influence of the polarization of light. In Table~\ref{magicwavfr2} magic wavelengths for both linear and circular polarization of light are given.

\section{\label{sec:tuneout}Tune-out wavelengths}
\begin{figure}[]
\includegraphics[width=82 mm,angle=0]{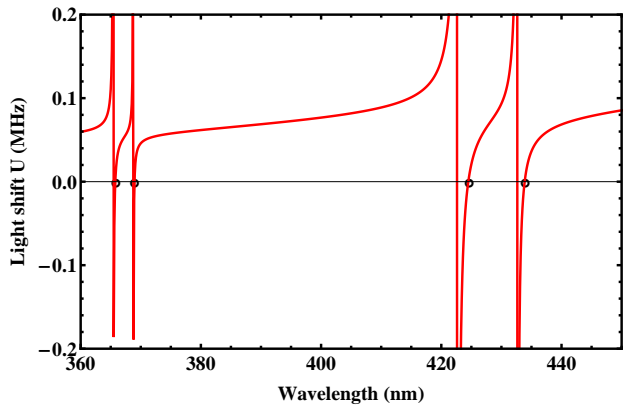}
\caption{\label{tuneout} (Color online) Tune-out wavelengths of the $^{212}$Fr isotope for the $7s~^{2}S_{1/2}$ ground state corresponding to the $7s~^{2}S_{1/2}-8p~^{2}P_{1/2,3/2}$ and $7s ^{2}S_{1/2}-9p~^{2}P_{1/2,3/2}$ transitions. Tune-out wavelength of the $7s~^{2}S_{1/2}-7p~^{2}P_{1/2,3/2}$ transitions is not shown. The position of the tune-out wavelengths is designated with black circles.}
\end{figure}

For an atomic state, tune-out wavelength is defined as the wavelength for which frequency dependent polarizability become zero. The details of the concept and advantages of tune-out wavelengths are discussed in~\cite{Leblanc2007,arora2011}. Tune-out wavelengths were calculated for stable alkali atoms~\cite{arora2011}. Here we present the calculated tune-wavelengths for the $7s-7p$, $7s-8p$ and $7s-9p$ lines of $^{212}$Fr and the $7s-7p$ transitions of $^{210}$Fr and $^{221}$Fr. The energy level values for the three isotopes are taken from~\cite{sansonetti2007} and the transition rates are the same, which are also taken from~\cite{sansonetti2007}. In Fig.~\ref{tuneout} tune-out wavelengths of the $7s-8p$ and $7s-9p$ lines for $^{212}$Fr isotope are shown and the values are given in Table~\ref{tuneoutfr}. In addition, for the $7s-7p$ line of the $^{210,212,221}$Fr isotopes tune-out wavelengths are calculated from the lifetimes of the $7p~^{2}P_{1/2,3/2}$ levels and they are also given in Table~\ref{tuneoutfr}. The uncertainty is 0.03~nm for the tune-out wavelengths, which is from the wavelengths of the transitions and in identifying the position.

\begin{table*}[htb]
\begin{center}
\caption{Magic wavelengths (vacuum) of francium atom for $7s~^2S_{1/2}-7p~^{2}P_{1/2,3/2}$ transitions. Wavelengths are given for both linear and circular polarization of the $m_{J}=1/2$ and $m_{J}=3/2$ states.}
\begin{tabular}{cccccccccc}
\hline \hline
 Transition                        &  Wavelength (nm)  &            & Transition  & & Wavelength (nm)  & &  & &\\
                                   \cline{2-3}                              \cline{5-8}          \\
                                   & $m_J=1/2$     &            &            & $m_J=1/2$     & & $m_J=3/2$ & \\ 
 \hline
  7s~$^{2}S_{1/2}$-7p~$^{2}P_{1/2}$ &  $\Delta m_J=0$ & $|\Delta m_J|=1$ & 7s~$^{2}S_{1/2}$-7p~$^{2}P_{3/2}$ & $\Delta m_J=0$ & $|\Delta m_J|=1$ & $\Delta m_J=0$ & $|\Delta m_J|=1$\\
                                   & 366.21      & 367.58      &                                   & 366.20       & 366.55              & 365.99       & 368.10 \\
                                   & 369.39      & 466.17      &                                   & 369.35       & 429.19              & 369.14       & 442.61\\
                                   & 426.67      & 495.07      &                                   & 426.51       & 537.36              & 425.63       & 508.44 \\                                                                          
                                   & 437.74      & 496.54      &                                   & 437.15       & 540.35              & 435.26       & 540.03 \\                              
                                   & 489.34      & 498.30      &                                   & 533.63       & 541.88              & 538.94       & 541.62 \\                               
                                   & 495.19      & 500.54      &                                   & 540.16       & 544.97              & 540.68       & 544.54 \\                              
                                   & 496.65      & 503.42      &                                   & 541.70       & 547.56              & 541.34       & 546.68 \\                              
                                   & 497.12      & 507.17      &                                   & 542.56       & 550.87              & 542.20       & 549.70 \\                               
                                   & 498.47      & 512.23      &                                   & 544.48       & 554.38              & 543.23       & 553.34 \\
                                   & 498.95      & 519.35      &                                   & 544.96       & 561.49              & 545.10       & 560.55 \\
                                   & 500.71      & 529.85     &                                   & 545.56       & 572.22               & 545.56       & 571.53 \\
                                   & 501.27      & 546.64     &                                   & 546.95       & 585.51               & 547.85       & 584.10 \\
                                   & 503.61      & 576.10     &                                   & 547.66       & 587.77               & 548.50       & 603.36 \\
                                   & 504.27      & 640.25     &                                   & 550.10       & 605.99               & 551.33       & 643.62 \\
                                   & 507.43      & 1116.2    &                                    & 551.13       & 645.60               & 552.29       & 730.64 \\
                                   & 508.16      &            &                                   & 553.78       & 649.57               & 555.04       &       \\
                                   & 512.41      &            &                                   & 555.51       & 729.76               & 561.93       &       \\ 
                                   & 513.46      &            &                                   & 560.74       & 731.43               & 564.08       &      \\
                                   & 519.39      &            &                                   & 561.82       & 1395.3               & 572.39       &  \\
                                   & 520.87      &            &                                   & 569.61       &                      & 573.60      &  \\
                                   & 529.47      &            &                                   & 572.13       &                      & 585.86       &  \\
                                   & 531.76      &            &                                   & 573.63       &                      & 587.75       & \\
                                   & 545.08        &            &                                 & 581.67       &                      & 606.66         & \\
                                   & 549.04        &            &                                 & 585.36       &                      & 610.20        & \\
                                   & 571.12      &            &                                   & 600.33       &                      & 645.95        & \\
                                   & 579.16      &            &                                   & 605.64       &                      & 649.51        & \\
                                   & 621.11      &            &                                   & 632.38       &                      & 729.73        & \\
                                   & 642.85      &            &                                   & 645.11       &                      & 731.77                &\\
                                   & 745.36      &            &                                   & 649.65       &                      & 783.35                &\\
                                   & 797.75     &            &                                    & 694.67       &                      & 967.19               & \\
                                   & 871.62      &            &                                   & 730.51             &                & 1111.7                & \\                                  
                                   &             &            &                                   & 731.32             &                    &                & \\                                  
                                   &             &            &                                   & 784.62             &                    &                & \\                                  
                                   &             &            &                                   & 853.93             &                    &                & \\                                  
                                   &             &            &                                   & 968.83             &                    &                & \\                                  
                                   &             &            &                                   & 1266.3             &                    &                & \\

 \hline \hline
\end{tabular}
\label{magicwavfr1}
\end{center}
\end{table*}

\begin{table}[htb]
\begin{center}
\caption{Magic wavelengths for the $7s~^{2}S_{1/2}-8s~^{2}S_{1/2}$ transition in francium atom. Wavelengths (vacuum) are given for both linear and circular polarization of light.}
\begin{tabular}{cccccccc}
\hline \hline
 Transition &  Wavelength (nm) & \\
  \hline
 7s~$^{2}S_{1/2}$-8s~$^{2}S_{1/2}$ & $\Delta m_J=0$ \\
                                   &   365.78   \\
                                   &   368.97   \\
                                   &   424.51   \\
                                   &   433.89   \\
                                   &   783.47   \\
                                   &   953.09   \\                                                                          
                                   &   960.15   \\                                                                          
                                   &  $|\Delta m_J|=1$ \\
                                   &   365.95   \\
                                   &   424.54   \\                           
                                   &   841.08   \\                               
                                   &   897.22   \\                              
                                   &   954.01   \\                              
                                  
 \hline \hline
\end{tabular}
\label{magicwavfr2}
\end{center}
\end{table}
    
\begin{table}[htb]
\begin{center}
\caption{Tune-out vacuum wavelengths for the $7s-7p$, $7s-8p$ and $7s-9p$ doublet lines of three francium isotopes. The wavelength values with an asterik(*) are from the experimental lifetimes of the $7p~^2P_{1/2,3/2}$ states.}
\begin{tabular}{cccc}
\hline \hline
 Isotope & Transitions &  Wavelength (nm) & \\
  \hline
 $^{210}$Fr&$7s~^{2}S_{1/2}$-$7p~^{2}P_{1/2,3/2}$ & 782.87, 781.56$^{*}$ \\
 $^{212}$Fr&$7s~^{2}S_{1/2}$-$7p~^{2}P_{1/2,3/2}$ & 782.96, 781.65$^{*}$ \\
          &$7s~^{2}S_{1/2}$-$8p~^{2}P_{1/2,3/2}$  & 424.49  \\
          &$7s~^{2}S_{1/2}$-$8p~^{2}P_{1/2}$      & 433.87 \\
          &$7s~^{2}S_{1/2}$-$9p~^{2}P_{1/2,3/2}$  & 365.76 \\
          &$7s~^{2}S_{1/2}$-$9p~^{2}P_{1/2}$      & 368.96  \\                                 
 $^{221}$Fr&$7s~^{2}S_{1/2}$-$7p~^{2}P_{1/2,3/2}$ & 782.94, 781.63$^{*}$    \\
 \hline \hline
\end{tabular}
\label{tuneoutfr}
\end{center}
\end{table}

\section{Conclusion}
In conclusion, we have calculated the light shifts of the $7s~^{2}S_{1/2}$, $7p~^{2}P_{1/2,3/2}$, and $8s~^{2}S_{1/2}$ levels in francium, which is being pursued for fundamental symmetry investigations using laser cooling and trapping methods. The light shifts are calculated from the available data of energy levels and transition rates. The magic wavelengths for the state insensitive optical dipole trapping of atoms are identified for the D1 and D2 transitions of francium from the light shifts. Magic wavelengths in the ultraviolet, visible and near infrared region are identified corresponding to both blue and red detuning of frequency with respect to the laser cooling and trapping transitions. In particular, magic wavelengths between 600-700~nm and 700-1000~nm are convenient for the implementation of the optical dipole trap as lasers with sufficient power are available in these wavelengths range. Also, magic wavelengths for the forbidden $7s~^{2}S_{1/2}-8s~^{2}S_{1/2}$ transition are obtained from the light shifts of the $7s~^{2}S_{1/2}$ level and the $8s~^{2}S_{1/2}$ level. Furthermore, tune-out wavelengths where the ground state ac-polarizability of the francium atom vanishes is reported for three of the isotopes.
 
\begin{acknowledgments}
This work is carried out under the research programme of JSPS KAKENHI (26220705). Y.S. is supported by JSPS and INSA Bilateral Joint Research Project. One of the authors U.D. acknowledges kind help from  Lotje Wansbeek, The Netherlands and Arun.
K. Thazathveetil, Department of Chemistry, North-Western
University, Evanston, USA.
\end{acknowledgments}

\bibliography{Frreferences}

\end{document}